# Automatic Characterization of Mid-latitude Multiple Ionospheric Plasma Structures from All-sky Airglow Images using Deep Learning Technique

Jeevan Upadhyaya[1], Satarupa Chakrabarti[2], Rahul Rathi[3], Virendra Yadav[4], Dipjyoti Patgiri[1], Gaurav Dixit[5,6], M.V. Sunil Krishna[1,7], Sumanta Sarkhel[*1,7]

[1]Department of Physics, Indian Institute of Technology Roorkee, Roorkee – 247667, Uttarakhand, India.
[2]School of Electrical Engineering and Computer Science, KTH Royal Institute of Technology, 114 28 Stockholm, Sweden
[3]Physics Department, Lancaster University, Lancaster, United Kingdom
[4]Aryabhatta Research Institute of Observational Sciences, Nainital – 263001, Uttarakhand, India
[5]Department of Management Studies, Indian Institute of Technology Roorkee, Roorkee – 247667, Uttarakhand, India
[6]Mehta Family School of Data Science and Artificial Intelligence, Indian Institute of Technology Roorkee, Roorkee 247667, Uttarakhand, India
[7]Centre for Space Science and Technology, Indian Institute of Technology Roorkee, Roorkee –247667, Uttarakhand, India

[*]Sumanta Sarkhel (sarkhel@ph.iitr.ac.in)

**Abstract**

The F-region ionospheric plasma structures are propagating high and/or low electron density regions in the Earth's ionosphere. These plasma structures can be observed using ground based all-sky airglow imagers which can capture faint airglow emissions originating from the F-region of ionosphere. This study introduces a novel automatic method for determining the propagation parameters (horizontal velocity and orientation) of these multiple ionospheric plasma structures observed in $O(^1D)$ 630.0 nm all-sky airglow images from Hanle, India located in the mid-latitude region. We have used a deep learning-based segmentation model called YOLOv8 (You Only Look Once) to localize and BoT-SORT tracker to track individual mid-latitude ionospheric plasma structures. Three different automatic algorithms are used to characterize the observed plasma structures utilizing the segmented outputs from the YOLO model. Finally, an additional quality control step is introduced that filters the results from the three automatic algorithms and generates a flag to retain the most reliable estimate. The results of the proposed fully automated pipeline are systematically compared with a previously developed semi-automatic approach to assess the estimation efficacy. The automatic technique developed in this study is particularly valuable for all-sky airglow imaging systems having large datasets, where manual intervention or semi-automatic analysis is impractical.

**Keywords:** All-sky Airglow Imager, Mid-latitude Ionospheric Plasma Structures, Deep Learning, YOLO, Automatic Characterization.

**Key Points:**

1. Fully automatic method is developed to characterize ionospheric plasma structures using multiple approaches to reduce manual intervention.
2. A combination of YOLOv8 and BoT-SORT tracker is used to localize and track multiple plasma structures.
3. Quality control step is introduced to remove unreliable results thus making the proposed method suitable for large datasets.

**Plain Language Summary**

Ionospheric plasma structures are localized enhanced/depleted regions of plasma in the Earth's ionosphere that can affect radio wave communication and navigation system. Different ground and satellite-based instruments are used to detect and examine these structures. All-sky airglow imagers, that capture faint emission coming from the Earth's ionosphere, are frequently used to study plasma structures. Processes such as electrodynamical interaction, merging, distortion, and dissipation can modify the horizontal velocity and orientation of plasma structures. Therefore, reliable estimation of these parameters is crucial for understanding their underlying electrodynamics. This study presents a fully automated framework to estimate the propagation parameters such as horizontal velocity and orientation of mid-latitude ionospheric plasma structures. A deep learning-based YOLO model is used to segment and track plasma structures, followed by three automatic algorithms for their characterization. A quality-control step filters and combines these results to retain only reliable estimate. The results are also compared against an existing semi-automatic method and is well suited for large datasets.

# 1. Introduction

The mid-latitude ionosphere hosts a variety of plasma irregularities that significantly influence ionospheric electrodynamics and radio wave propagation. Frequently observed plasma irregularities/structures at mid-latitudes are medium-scale traveling ionospheric disturbances (MSTIDs) and mid-latitude field-aligned plasma depletions (Garcia et al., 2000; Hocke and Schlegel, 1996; Otsuka, 2021; Shiokawa et al., 2003 and references therein). Over the last few decades, these structures have been extensively studied using different ground and spaced-based instruments such as ionosonde, airglow imager, Global Navigation Satellite System (GNSS), and radar (Candido et al., 2008; Huang et al., 2018, 2021; Liu et al., 2021; Maeda & Heki, 2015). Ground-based airglow imagers are well suited for investigating these structures which appear as localized regions of enhanced and depleted intensity in $O(^1D)$ 630.0 nm airglow images (Hozumi et al., 2025; Makela, 2006; Otsuka et al., 2004; Patgiri et al., 2024b; Rathi et al., 2025; Shiokawa et al., 2003; Sivakandan et al., 2021; Tsuchiya et al., 2020). Previous studies have shown that the propagative characteristics of ionospheric plasma structures, particularly their horizontal propagation velocity and orientation, can exhibit drastic changes arising due to interactions between different plasma structures (Otsuka et al., 2012; Rathi et al., 2022; Sun et al., 2015; Wu et al., 2021). Therefore, reliable estimation of propagation parameters is crucial for understanding the underlying physical processes.

In recent years, different methodologies have been proposed to estimate the propagation parameters of plasma structures (Chakrabarti et al., 2025; Lai et al., 2023; Yadav et al., 2021b). A semi-automatic approach developed by Yadav et al. (2021b) and Patgiri et al. (2024c) has been employed to characterize multiple MSTID fronts. However, the reliance on semi-automatic workflow demands manual intervention incurring subjectivity bias and human error, making the analysis time-consuming, and poorly suited for datasets spanning over many years. On the other hand, methods such as 3D spectral analysis and autocorrelation do not involve manual intervention and can determine horizontal velocity, wave period, and propagation direction (Naito et al., 2022; Takeo et al., 2017; Tsuboi et al., 2023). However, these methods are unable to estimate the propagative parameters and orientation of individual plasma bands coexisting in a single event.

Apart from existing semi-automatic and algorithmic approaches, deep learning techniques have been increasingly explored as a suitable alternative for automatic characterization of different types of ionospheric irregularities. Deep learning methodologies have demonstrated substantial potential in handling large-scale datasets through automated detection, classification, and localization, thus reducing human intervention while improving consistency and reproducibility (Guerra et al., 2025; Kapil & Seemala, 2024; Mutasov et al., 2025). Authors have applied deep learning techniques to the all-sky airglow imager datasets for the automatic detection and characterization of equatorial plasma bubbles, a commonly observed low-latitude plasma structure (Okoh et al., 2025; Zhong et al., 2025). Lai et al. (2023) developed CNN-based classification and Fast R-CNN–based model to characterize and analyze the statistical occurrence of MSTIDs. Although Chakrabarti et al. (2025) developed an automated framework to characterize mid-latitude plasma structures from all sky airglow images, however this method is limited to only single band.

Recent deep learning-based studies have brought forward the YOLO architecture to be a suitable choice as it is powered by a single forward pass regression framework, thus reducing inference time and making accurate instance segmentation. Zhang et al. (2026) applied a variant of this model (YOLO-LessHead model) to Limb and Disk (GOLD) 135.6 nm nightglow dataset for automatic labeling. Furthermore, Le et al. (2025) employed a YOLO-based framework on detrended TEC (dTEC) data for detection and characterization of mid-latitude ionospheric plasma irregularities. As discussed, prior studies have implemented YOLO-based frameworks for detecting and estimating propagation characteristics of ionospheric plasma structures. However, to the best of our knowledge, YOLO-based instance segmentation and continuous tracking of individual plasma structures from all-sky airglow observations has not been reported in the literature so far.

In the present study, we have developed a fully automated pipeline for localization, tracking and parameterization of individual mid-latitude ionospheric plasma structures using all-sky airglow images. Apart from using the deep learning framework to localize and track the plasma structures, we have used three standalone approaches to estimate the propagation parameters. A major contribution of this work is the inclusion of a quality filter that helps in processing the output from different approaches and retain only the most reliable estimate. The paper is structured as follows. Instance segmentation and tracking, automatic calculation of parameters and quality filtering is explained in Section 2 along with the dataset description. Section 3 presents the evaluation of the results. A detailed discussion is included in Section 4, followed by conclusion in Section 5.

## 2. Dataset and Methodology

### 2.1 Data Description

The present study utilizes data from a multi-wavelength all-sky airglow imager installed at Hanle, India (32.77°N, 78.97°E; Mlat. ~24.1°N). The imager captures $O(^1S)$ 557.7 and $O(^1D)$ 630.0 nm airglow emissions with a field-of-view (FOV) of ~140°. These images undergo a sequence of preprocessing steps which include geospatial calibration (computation of latitude and longitude corresponding to each pixel), noise removal, and geometric unwarping. A detailed description of these preprocessing steps is discussed in Mondal et al. (2019). As the present study revolves around characterization of F-region plasma irregularities, therefore we have only utilized 630.0 nm airglow images.

The processed unwarped images used are single-channel grayscale images with a resolution of 511×511 pixels. The complete dataset consists of 1768 images spanning over 7 years (2018 to 2025). In the present analysis, we only considered images that contain the plasma structures using a previously developed stand-alone CNN classifier (Chakrabarti et al., 2024). However, there are instances where these structures appear faint with diffused boundaries. To enhance their appearance and visual contrast, intensity scaling is performed. Our dataset consists of plasma structures that exhibit substantial morphological diversity with single or multiple coexisting bands within the FOV (Chakrabarti et al., 2024, 2025). Figure 1 presents few instances of plasma structures with different morphology and band count. As the present work revolves around localization of the plasma structures, we require the mask data or the ground truth along with our original dataset. In this work we have used polygonal boundary for generation of ground truth as traditional bounding box will fail to capture the spatial extent of individual plasma structure. The polygonal boundary annotation is performed using the VGG Image Annotator (VIA) software (Dutta & Zisserman, 2019).

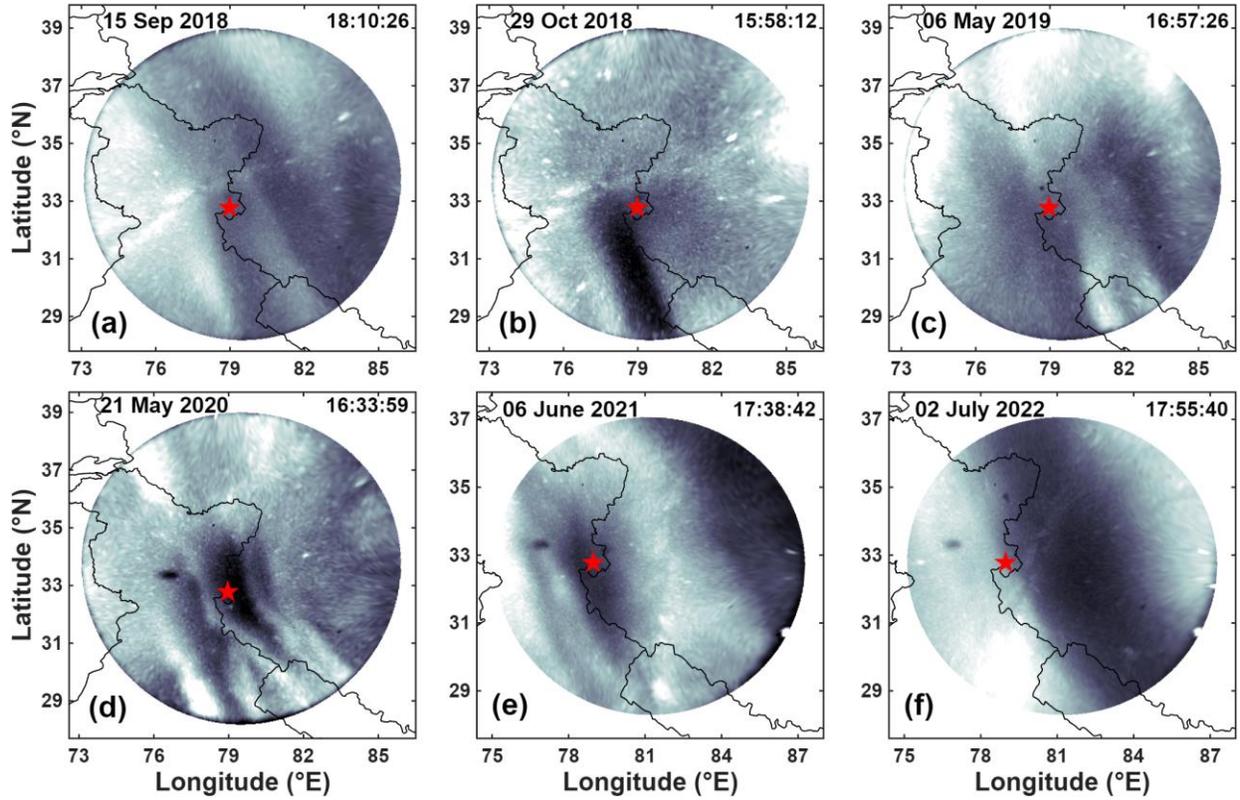

**Figure 1.** Plasma structures with different morphologies captured by allsky airglow imager at 630.0 nm wavelength. Red star indicates the location of the imager.

## 2.2. Proposed Methodology

The present study explores a combination of steps executed in a pipeline. The first step is the instance segmentation and tracking that utilizes the combination of YOLOv8 and BoTSORT architectures. The second step consists of automatic calculation of propagative parameters. The final step is the quality control to remove outlier values and enhance the efficacy of our proposed approach. The following subsections cover the detailed description of the proposed methodology. The algorithm of the deep learning framework, along with the working principle of the three automatic methods and quality filter is mentioned as pseudo codes in the Supplementary file.

### 2.2.1 Step1: Instance Segmentation and Tracking

**(a) Architecture of YOLOv8-seg and BoT-SORT**

The aim of the present study is to localize and track each plasma structure for estimating the propagative parameters. To achieve this objective, we have used the combination of YOLO and BoT-SORT models. YOLO is effective in object detection, and we have tested its different versions before finalizing the YOLOv8-seg architecture for instance segmentation of the plasma structures (Jocher et al., 2023; Redmon et al., 2016). YOLOv8-seg model produced considerably more accurate segmentation masks compared to its other versions and demonstrated better stability metrics. The YOLOv8-seg provides class labels, bounding boxes, and masks along with confidence scores frame by frame. However, inter-frame dependency cannot be processed by YOLO. In order to track a particular object in an image sequence, unique tracking identifier (Track ID) is required to be appended with the output of YOLO. Therefore, in the current study a modified BoT-SORT (Bag of Tricks - Simple Online Realtime Tracker) tracking algorithm is implemented to the output of the YOLOv8 detection stage (Aharon et al., 2022). Figure 2 presents the architecture of the YOLOv8-seg and BoT-SORT models.

As shown in Figure 2, the architecture of YOLOv8-seg consists of three primary parts- Backbone, Neck, and Head. Backbone, a series of convolutional layers (P1 to P5) arranged in a shallow to deep bottom-up fashion, is responsible for extracting feature maps of different dimensions from the input images. C2f Module (Cross-Stage Partial bottleneck with two convolutions) in the backbone effectively merges concatenated features, minimizes redundant computation, and strengthens feature representation by reusing inter-channel information. The Neck is responsible for combining the extracted features before passing to the detection Head. This feature fusion enables robust detection of objects across a wide range of scales, which in our case corresponds to the varying sizes of mid-latitude ionospheric irregularities. Finally, the Head infers the bounding boxes and their corresponding masks with a confidence score. The Head consists of three parallel detection units corresponding to different spatial resolutions for detecting objects. In each detection unit there are 2 parallel branches - a classification branch which predicts the class probability and a regression branch which predicts the bounding box coordinates. An additional segmentation head is present that predicts a fixed set of global prototype masks, which are linearly combined with the mask coefficients produced by the detection head to generate masks required for instance segmentation of the plasma structures.

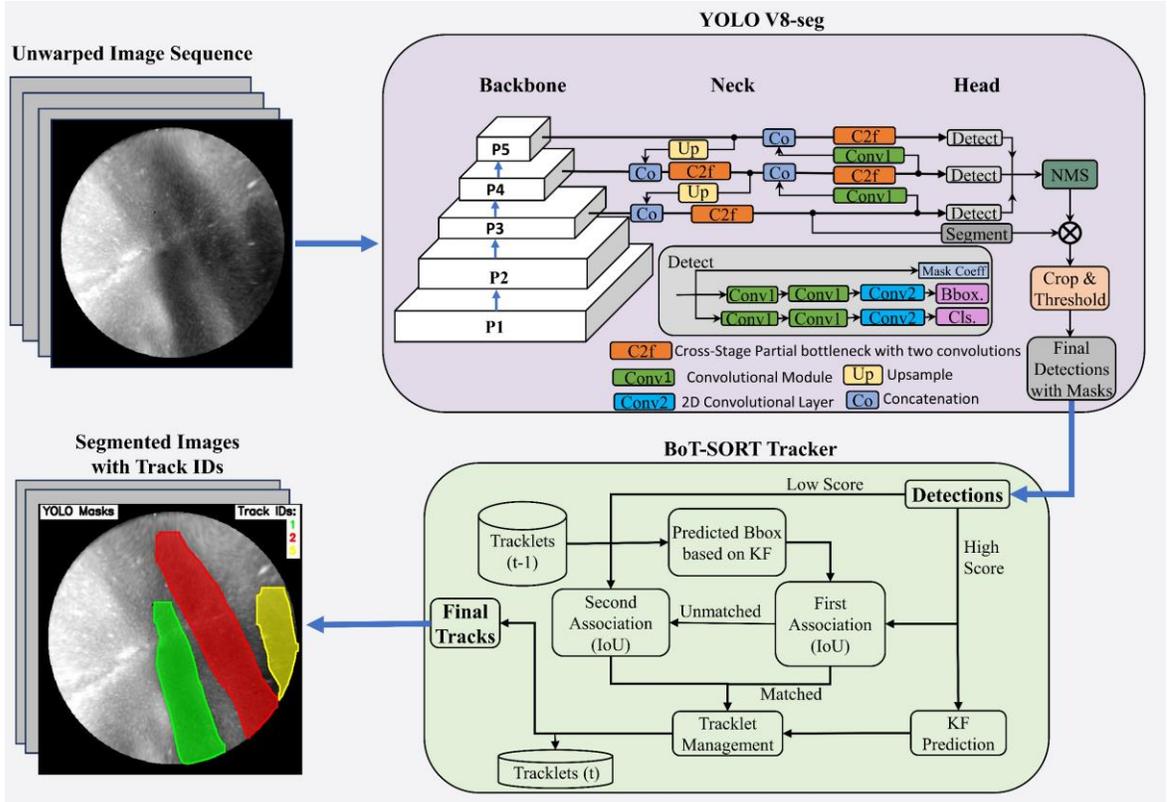

**Figure 2.** Flowchart of the proposed automatic pipeline for automatic segmentation of ionospheric plasma structures.

The masks generated by YOLOv8-seg lack Track IDs essential for identifying each plasma structure across the image sequence. Therefore, BoT-SORT tracker is introduced into the pipeline as shown in the Figure 2. This tracker operates by initializing a Kalman Filter for each identified structure (Aharon et al., 2022). Since the images used in the present study consist of non-rigid objects (plasma structures) and are captured by a stationary airglow imager, therefore, we disabled Re-Identification (ReID) network and Global Motion Compensation (GMC). Generation of Track IDs for each plasma structure and maintaining their accurate assignment across the image sequence is achieved using the Kalman Filter and IoU (Intersection over Union). In order to mitigate the loss of frequently observed faint structures, the detection thresholds are relaxed. This configuration ensures that even weak detections are kept as a valid candidate as long as they have sufficient spatial overlap with predicted location by Kalman Filter.

**(b) Training, Validation, and Hyperparameter Configuration of YOLOv8-seg**

In order to execute the YOLOv8-seg model for localization of plasma structures, we have divided our dataset into training and validation subset while keeping a separate dataset for testing.

The splitting of original dataset into training and validation is done using date-wise partitioning strategy instead of randomized partitioning. The reason behind using date-wise partitioning is that our dataset has temporal dependency and standard random splitting approach will make the model memorize morphological features which may artificially inflate the performance metrics (Sedlak et al., 2023; Zhong et al., 2025). The training set consists of 1253 images from June 2018 to May 2025. Similarly, a validation set of 251 images from January 2019 to June 2025 is selected which is mutually exclusive with training dataset. In addition, data augmentation is also used to increase diversity by applying random transformations like small rotations and translation and along with random fine brightness variations (Shorten & Khoshgoftaar, 2019).

After splitting the dataset, the model is initially trained for 300 epochs with a batch size of 8 and pre-trained COCO (common object in context) weights. Like all deep learning models, YOLO also has a set of hyperparameters that can be tuned according to the dataset. As our aim is to make the model converge with minimum error rate, we have tuned few specific hyperparameters such as – initial learning rate(lr0), final learning rate(lrf), batch size, number of epochs, optimizer, weight decay and momentum. After training and validation, the best model is automatically saved by the Ultralytics framework. Figure 3 presents the final training and validation loss curve along with the performance metrics. The orange curves show the reduction in training and validation losses categorized into Box, Seg (segmentation), Cls (classification), and Dfl (distribution focal) losses. Whereas the blue curves illustrate the improvement in Precision, Recall, and Mean Average Precision (mAP) for the Mask and Box metrics. The training and validation of the model are conducted on a single NVIDIA-RTX-3060 GPU with the total training time spanning over nearly 8 hours.

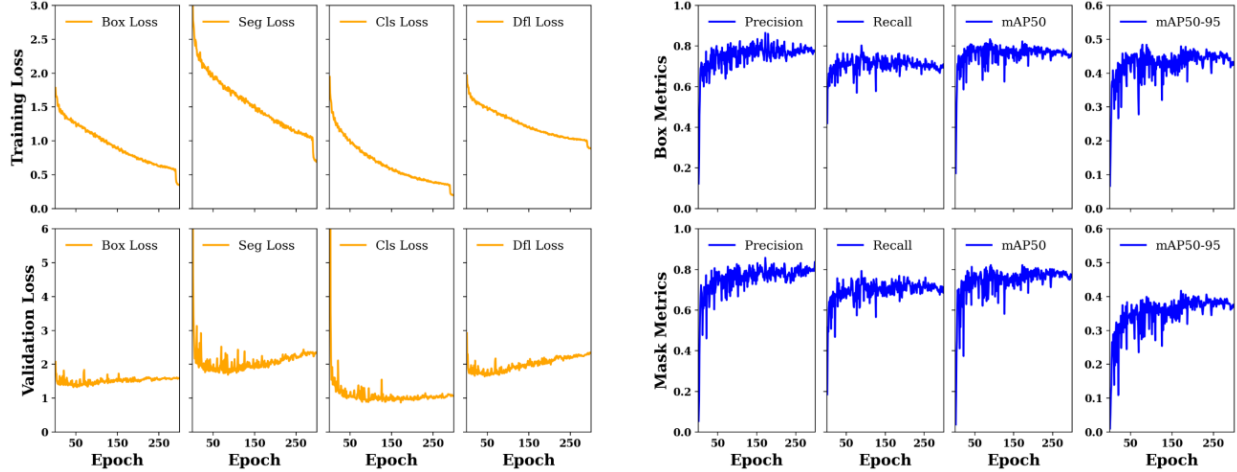

**Figure 3**. Training and validation loss curves and performance metrics of YOLOv8-seg model trained on allsky airglow unwarped images.

### 2.2.2. Step2: Automatic Calculation of Propagative Parameters

After segmentation using YOLOv8-seg and tracking the plasma structures by the BoT-SORT tracker, the model output along with the unwarped original images are fed into three distinct automated algorithms to estimate propagation parameters. Minima method, Maximum Normalized Cross-Correlation (MNCC), and Optical Flow are the three algorithms used in the present study. They are schematically illustrated in Figure 4.

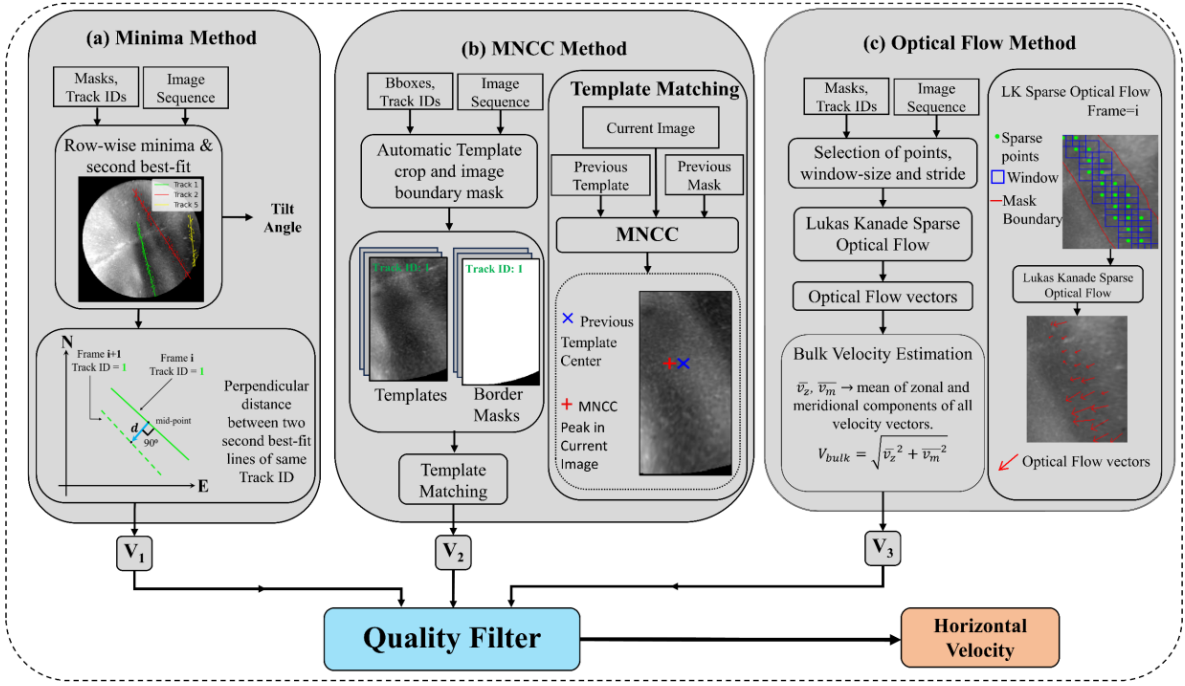

**Figure 4.** Schematic representation of the three algorithms. (a) Intensity minima approach, (b) Masked Normalized Cross-Correlation (MNCC), and (c) Optical flow to estimate the parameters.

### (a) Minima Method

The Minima Method utilizes the segmented masks and track IDs along with the unwarped image sequence. Drawing upon previous research by Chakrabarti et al. (2025), the algorithm works by drawing the best-fit line of row-wise minimum intensity associated with the plasma structures. This minimum intensity line nearly aligns with the plasma structure and provide the orientation/tilt angle of the structure. The perpendicular distance between the minima lines of two successive frames with same Track ID is used to determine the horizontal propagation velocity ($V_{Minima}$) of a particular plasma structure. In the previous work (Chakrabarti et al., 2025) the minima method was used only on single band plasma structures to calculate the propagative parameters. However, in the present work with the help of Track IDs we are able to mitigate the previous limitation and find the parameters of single as well as multiple plasma structures. Uncertainties associated with each calculation of the tilt angle and horizontal propagation velocity were estimated using the standard error obtained from the spatial spread of the intensity minima points. The estimated mean uncertainties are around 0.4° in tilt angle and ~10 m/s in horizontal propagation velocity is observed for the test dataset.

**(b) MNCC Method**

The second technique to estimate the horizontal propagation velocity is MNCC (Maximum Normalized Cross-Correlation) (Padfield, 2012). MNCC tracks the motion of the entire 2-dimensional pattern of the structure in the images. This method uses the concept of template matching, where a template of the image containing the target structure is taken from the previous frame and is moved over the subsequent frame to calculate the correlation matrix. In the present study, the template is cropped from previous frame using the rectangular bounding boxes obtained after applying the segmentation algorithm. MNCC generates a correlation map with pixel values ranging from -1 to +1 (+1 indicating perfect linear match). The location of the maximum correlation value (peak) corresponds to the new position of the plasma structure in the next frame. The pixel displacement is then calculated by measuring the shift between the previous template center and the MNCC peak in current frame. This displacement is then converted into geographical coordinates to estimate the horizontal velocity ($V_{MNCC}$) of the plasma structure. The uncertainty in the velocity is quantified using the peak to second peak ratio of correlation matrix (Charonko & Vlachos, 2013). The mean uncertainty of the estimated horizontal propagation velocity in the test dataset was approximately 11 m/s.

**(c) Optical Flow**

We have also implemented the Optical Flow (OF) approach, which is widely used in object tracking (Kale et al., 2015; Moing et al., 2024; Leonida et al., 2022), using the Lucas-Kanade sparse optical flow technique (Lucas & Kanade, 1981). Optical flow estimates displacement by analyzing the apparent motion of brightness patterns of pixels between consecutive image frames and assumes that pixels within a small window move coherently. The output of the method is a motion vector referred as 'flow vector', which is the mathematical representation of magnitude and direction of pixel movements in two consecutive images. Optical flow requires distinct corners or sharp features across the entire image (Beauchemin & Barron, 1995; Sun et al., 2014), but due to lack of sharp features in the plasma structures our approach employs a re-initialization strategy. The re-initialization strategy ensures that the velocity estimate for a given frame pair is derived from points that are specifically inside the plasma structures by utilizing the masks and track IDs provided by the segmentation algorithm. Finally, we selected the combination of window size and stride that minimized the directional variance. Once all the flow vectors are obtained, the displacements in pixel coordinates are transformed to geographic coordinates to calculate horizontal propagation velocities. The global motion of the plasma structure is represented by the bulk velocity; $V_{Optical\ Flow} = \sqrt{\overline{v_z}^2 + \overline{v_m}^2}$ that is derived by averaging zonal ($v_z$) and meridional ($v_m$) velocity components of all tracking points. The uncertainty in the bulk velocity is estimated using Standard Error of Mean (SEM) of the velocity magnitudes derived from all tracking points where the horizontal propagation velocity exhibited a mean uncertainty of ~8 m/s in the test dataset.

### 2.2.3 Step3: Quality Filter

The three methods described above independently estimate horizontal propagation velocities denoted as $V_{Minima}$, $V_{MNCC}$, and $V_{Optical\ Flow}$. As we have used three different methods, therefore there is a possibility that the estimated horizontal velocities show larger spread or give outlier values. These outlier points can be due to insufficient texture in the MNCC method, high brightness variation in the optical flow method or large structural spread in the minima method. Thus, a simple mean of such estimates leads to erroneous results. In order to obtain a reliable estimate of horizontal velocity from the three methods, we have introduced the Quality Filter that rejects these outlier values before calculating the final filtered velocity. First step involves the standard deviation approach to select the velocity values which are very close. To quantify this closeness, we have applied a very strict criterion of standard deviation less than 4 m/s (which is less than 15% of velocity values of all considered events). When the spread among the three velocity estimates is within this standard deviation threshold, the values are considered to be consistent and the mean gives the final horizontal velocity. The values which fail to satisfy the strict criterion proceed to the next step of outlier rejection.

This outlier rejection process is based on the Dixon's Q-Test (Dean & Dixon, 1951). It begins with arranging the velocities in ascending order ($v_1 \leq v_2 \leq v_3$) and later estimates the potential outlier (either lowest or highest value) based on the Q-value which is defined as –

$$Q_{low} = \frac{v_2 - v_1}{v_3 - v_1}, \quad Q_{high} = \frac{v_3 - v_2}{v_3 - v_1} \tag{1}$$

These calculated Q-values are compared against a critical threshold ($Q_{crit}$) depending on the confidence level (Rorabacher, 1991). In the current analysis, the critical value of 0.7403 is adopted which represents the mean Q-value of the test dataset. Additionally, the quality filter assigns a flag value to indicate the reliability of the final filtered horizontal velocity. A flag value of 1 denotes a good quality measure with low variability among the estimates. Whereas a value of 0.5 indicates moderate reliability, with moderate spread between the estimates and the result should be interpreted with caution. However, a flag value of 0 indicates that all estimates have been rejected.

## 3. Results

The performance of the YOLOv8-seg and BoT-SORT is first evaluated by examining its ability to perform instance-level segmentation and tracking of mid-latitude ionospheric plasma structures. The outputs for the test data confirm that the model is able to successfully isolate individual plasma structures and assign consistent track IDs across successive frames. The representative tracked instance masks for the test data are provided as movie (s1) in the Supplementary file. The bounding boxes, masks, and track IDs generated by YOLO serve as inputs to the subsequent automatic parameter estimation algorithms that enable structure-wise characterization without manual intervention.

The characterization results of the test dataset obtained from the automated methods are compared with semi-automatic results reported in previous studies (Chakrabarti et al., 2025; Patgiri et al., 2024c; Rathi et al., 2021, 2022; Yadav et al., 2021a; Yadav et al., 2021b). The test dataset consisted of 10 distinct events (single as well as multiple bands) with 19 plasma structures and have 120 values which were not included in the training dataset. These values represent the magnitude of propagative parameters (horizontal velocity, tilt angle, and direction of propagation). Figure 5 presents a quantitative comparison of these propagative parameters with semi-automatic method.

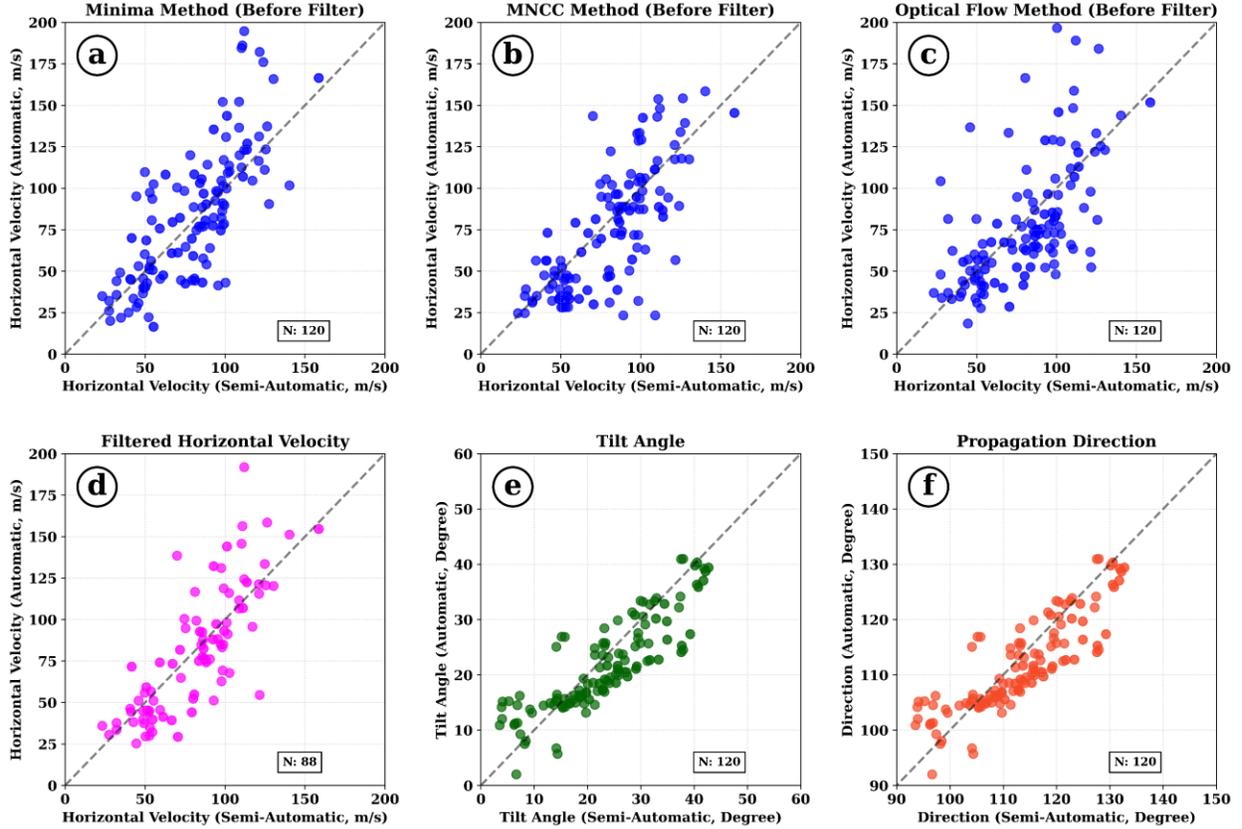

**Figure 5.** Scatter plot comparison of propagation parameters obtained via different algorithms with semi-automatic method. (a – c) Horizontal velocity obtained from Minima, MNCC, and Optical Flow methods. (d) Filtered horizontal velocity. (e & f) Tilt angle and propagation direction.

Figures 5a–c show the comparison of the magnitudes of horizontal velocity obtained using three automatic methods with the semi-automatic method (Yadav et al., 2021b; Patgiri et al., 2024c). This semi-automatic method requires manual intervention, where a straight line is drawn by visual inspection at the edge of the plasma structure, in determining the propagative parameters. In the figure, scatter points close to the diagonal line ($y = x$ line) denote perfect agreement. Figure 5d shows the filtered horizontal velocity after applying the quality filter. Although, a fraction of points (~26%) gets filtered, with the inclusion of flag the quality of the final estimate is greatly improved. Furthermore, the final filtered estimates of tilt angle and propagation directions are also illustrated in Figures 5e–f.

While the proposed methodology worked well for majority of the events in the test dataset, for a couple of events its performance degraded due to the presence of rapidly evolving/decaying plasma structures. This is presented with two such examples (15 September 2018 and 06 May 2019) in Figure 6. The dashed line with square markers shows the velocity from the proposed automatic method after quality filtering, while the solid line with circle markers represents the velocity obtained from the semi-automatic method. The points which are rejected after passing through the Quality Filter for both events are denoted as '×'. It is interesting to note that the rejected points in both the events are not significantly higher. In the case of 15 September 2018 event (Figures 6a&b), the estimates from both automatic and semi-automatic methods corroborate with each other. However, it is worth noting that in the 06 May 2019 event (Figures 6c&d), the spread of estimates from the three methods is higher due to changing morphology of the plasma structures in the presence of electrodynamical interaction. All the estimated values for these two types of events along with the flags are presented in Table 1.

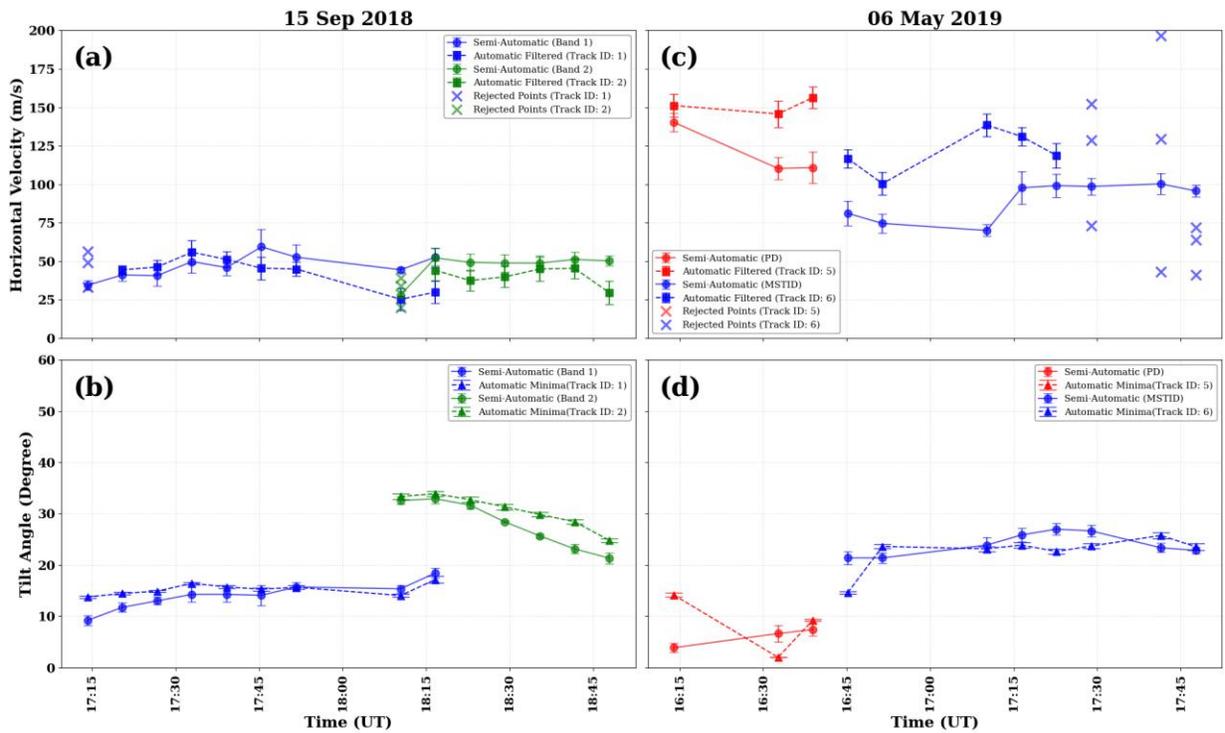

**Figure 6.** Comparison of temporal evolution of propagation parameters (horizontal velocity and tilt angle) along with their uncertainty estimates observed on 15 September 2018 and 06 May 2019.

**Table 1.** Final filtered velocity along with their respective quality flags

| Event Date | Time (UT) | Track ID | Velocity Minima Method (m/s) | Velocity MNCC Method (m/s) | Velocity Optical Flow Method (m/s) | Velocity Final (m/s) | Quality Flag |
|---|---|---|---|---|---|---|---|
| 15-Sep-2018 | 17:14:11 | 1 | 49.17 | 56.43 | 33.26 | - | 0 |
| | 17:20:26 | 1 | 44.85 | 56.43 | 44.03 | 44.44 | 1 |
| | 17:26:41 | 1 | 45.32 | 56.43 | 36.92 | 46.22 | 0.5 |
| | 17:32:56 | 1 | 60.22 | 56.43 | 51.03 | 55.89 | 1 |
| | 17:39:11 | 1 | 53.07 | 50.17 | 50.26 | 51.17 | 1 |
| | 17:45:26 | 1 | 45.38 | 45.60 | 55.11 | 45.49 | 1 |
| | 17:51:41 | 1 | 52.47 | 45.60 | 36.83 | 44.97 | 0.5 |
| | 18:10:26 | 1 | 95.14 | 32.15 | 18.46 | 25.30 | 0.5 |
| | 18:16:41 | 1 | 51.05 | 32.15 | 27.69 | 29.92 | 1 |
| | 18:10:26 | 2 | 20.05 | 39.16 | 33.88 | - | 0 |
| | 18:16:41 | 2 | 22.27 | 44.39 | 43.78 | 44.08 | 1 |
| | 18:22:56 | 2 | 39.65 | 39.16 | 33.64 | 37.48 | 1 |
| | 18:29:11 | 2 | 40.25 | 33.29 | 46.10 | 39.88 | 1 |
| | 18:35:26 | 2 | 45.53 | 33.29 | 44.45 | 44.99 | 1 |
| | 18:41:41 | 2 | 42.89 | 28.21 | 47.96 | 45.42 | 1 |
| | 18:47:56 | 2 | 40.50 | 28.21 | 30.89 | 29.55 | 1 |
| 06-May-2019 | 16:13:56 | 5 | 101.68 | 158.44 | 143.89 | 151.17 | 0.5 |
| | 16:32:40 | 5 | 112.57 | 143.15 | 148.30 | 145.72 | 1 |
| | 16:38:56 | 5 | 186.09 | 153.82 | 158.81 | 156.32 | 1 |
| | 16:45:11 | 6 | 44.29 | 122.29 | 111.13 | 116.71 | 0.5 |
| | 16:51:26 | 6 | 98.41 | 102.50 | 81.21 | 100.45 | 1 |
| | 17:10:10 | 6 | 100.48 | 143.53 | 133.50 | 138.51 | 0.5 |
| | 17:16:25 | 6 | 104.59 | 133.01 | 129.13 | 131.07 | 1 |
| | 17:22:41 | 6 | 117.06 | 133.36 | 105.88 | 118.77 | 0.5 |
| | 17:28:56 | 6 | 152.09 | 128.59 | 73.16 | - | 0 |
| | 17:41:26 | 6 | 43.16 | 129.29 | 196.60 | - | 0 |
| | 17:47:41 | 6 | 41.37 | 71.87 | 63.85 | - | 0 |

## 4. Discussion

This study explores the estimation of propagation parameters of mid-latitude ionospheric plasma structures using a combination deep learning technique and multi-method approach. Earlier studies primarily relied on semi-automatic tracking, spectral analysis, autocorrelation as well as keogram based methods to calculate propagation velocity and orientation (Otsuka et al., 2021; Sivakandan et al., 2019; Takeo et al., 2017; Yadav et al., 2021b). However, these approaches have limitations in structure-wise separation and not suitable for long-term datasets. More recent efforts have shifted toward deep learning-based frameworks that automate characterization and reduces manual intervention. In the last several years, YOLO-based detection and segmentation frameworks have demonstrated high efficiency in identifying ionospheric irregularities from satellite based (GNSS-TEC, GOLD) datasets (Le et al., 2025; Zhang et al., 2026). However, no prior study demonstrated the application of YOLO-based frameworks to ground based all-sky airglow imagery for structure-wise estimation of propagative parameters of multiple mid-latitude plasma structures.

The proposed fully automatic pipeline requires no manual intervention and provides a computationally efficient and less labor-intensive alternative to the existing semi-automatic approach. The semi-automatic approach tracks the edge of the plasma structure by manually defining a reference line, thus introducing a user-dependent bias. As semi-automatic method heavily relies on the edge tracking, therefore, when a plasma structure expands, dissipates, and translates simultaneously, the output of this method may result in inaccurate estimation of the parameters. Whereas the proposed pipeline requires no manual intervention and eliminates the human error from the analysis. In addition, the proposed automatic approach drastically reduces the processing time compared to the semi-automatic estimation of parameters for a single event. Therefore, when conducting statistical studies with large dataset, the semi-automatic approach becomes impractical. In addition, the semi-automatic method cannot estimate the data quality of the propagation parameters which can be very crucial while addressing the electrodynamics of plasma interactions in the ionosphere. Therefore, as described in the earlier reports, the authors had to perform the earlier semi-automatic method repeatedly for several times to find the spread in the estimation of the propagation parameters.

This shortcoming of the semi-automatic method is addressed in the proposed automatic method with a data quality indicator. It raises a flag based on the output of the quality filter and provides uncertainty estimate for each parameter. A flag value of 1 indicates a reliable estimate, 0.5 indicates an uncertain estimate that should be used with caution, and 0 indicates a poor-quality estimate that should be rejected. As we are shifting from semi-automatic to fully automatic approach, hence the quality filter and the flag play a key role in estimating the parameters. Although we have compared the results from our proposed automatic approach with the existing semi-automatic estimates, the latter cannot set as a standard reference due to its inherent issues described above. Therefore, the combination of quality filter and flag will help us in providing reliable estimates in future.

The criteria in the quality filter can be set to provide high quality (flag 1) or moderate quality (flag 0.5) depending on the usage of the application. A strict filtering criterion will yield high quality and consistent results but will consequently reduce the number of data points. One plausible reason for the rejection of points (as shown in Figures 5 & 6) is due to the morphology of the structures with evolving/decaying phases due to electrodynamical interaction among themselves (Yadav et al., 2021b). In stand-alone studies (especially in events with structures of changing morphologies due to interaction), this higher rejection of points can be a shortcoming of our approach as we may lose some inherent information about the structures. The statistics of the events (spanning over 7 years) containing plasma structures over the observation region shows that the number of events with interacting structures are significantly less (9.32 %). Therefore, in long-term statistical studies with large number of non-interacting events a moderate filtering criterion can be applied to compensate the information loss incurred due to the strict criterion. The proposed framework offers several advantages however, the performance can be further improved by adding more training and testing datasets from multiple all-sky airglow imagers to generalize the segmentation model and increase its robustness.

## 5. Conclusions

This study presents an automated pipeline for the localization and characterization of mid-latitude multi-band ionospheric plasma structures observed in O($^1$D) 630.0 nm images using deep learning technique. The highlights of the proposed framework are summarized as follows:

i. We have used the BoT-SORT tracker with the YOLOv8-seg architecture to segment and simultaneously track highly deformable mid-latitude ionospheric plasma structures.

ii. A multi-technique (Minima, MNCC, and Optical Flow) approach is applied on the output of segmentation architecture to determine the propagative parameters of the plasma structures.

iii. The main highlight of this work is the addition of a Quality Filter and Quality Flag. The Quality filter integrates the estimates from three independent methods and enforces strict criteria to eliminate outliers caused by the morphology of the structures and image artifacts. Along with the filter, the flag enables us to find the reliable estimates.

iv. The key contribution of this work is the ability to concurrently analyze and characterize individual bands from multiple coexisting plasma structures.

v. The proposed pipeline eliminates the need of manual intervention thus making it a preferred choice for analyzing large dataset of all-sky airglow images and conducting statistical studies.

**Acknowledgements:**

S. Sarkhel acknowledges the financial support from the Anusandhan National Research Foundation, Government of India (CRG/2021/002052). The support from the Indian Astronomical Observatory (operated by the Indian Institute of Astrophysics, Bengaluru, India), Hanle, Ladakh, India, for the day-to-day operation of the imager is duly acknowledged. J. Upadhyaya acknowledges the fellowship from the University Grants Commission, Government of India. S. Chakrabarti acknowledges the fellowship from the Digital Futures Mobility Program (Project no. 8317), KTH, Stockholm, Sweden. R. Rathi acknowledges the fellowship from NERC grant NE/W003090/1, Physics Department, Lancaster University, United Kingdom. D. Patgiri acknowledges the fellowship from the Ministry of Education, Government of India, for carrying out this research work. The work is also supported by the Ministry of Education, Government of India.

# References:

Aharon, N., Orfaig, R., & Bobrovsky, B.-Z. (2022, July 7). BoT-SORT: Robust Associations Multi-Pedestrian Tracking. arXiv. https://doi.org/10.48550/arXiv.2206.14651

Beauchemin, S. S., & Barron, J. L. (1995). The computation of optical flow. *ACM Comput. Surv.*, *27*(3), 433–466. https://doi.org/10.1145/212094.212141

Bhardwaj, A., Gupta, A., Ahmed, Q., Singh, A., Gupta, S., Sarkhel, S., et al. (2023). Signature of Y-forking in ionogram traces observed at low-mid latitude Indian station, New Delhi, during the earthquake events of 2020: ionosonde observations. *Frontiers in Astronomy and Space Sciences*, *10*. https://doi.org/10.3389/fspas.2023.1170288

Candido, C. M. N., Pimenta, A. A., Bittencourt, J. A., & Becker-Guedes, F. (2008). Statistical analysis of the occurrence of medium-scale traveling ionospheric disturbances over Brazilian low latitudes using OI 630.0 nm emission all-sky images. Geophysical Research Letters, 35(17).

Chakrabarti, S., Patgiri, D., Rathi, R., Dixit, G., Sunil Krishna, M. V., & Sarkhel, S. (2024). Optimizing a deep learning framework for accurate detection of the Earth's ionospheric plasma structures from all-sky airglow images. *Advances in Space Research*, *73*(12), 5990–6005. https://doi.org/10.1016/j.asr.2024.03.014

Chakrabarti, S., Upadhyaya, J., Rathi, R., Yadav, V., Patgiri, D., Dixit, G., et al. (2025). Automatic localization and characterization of mid-latitude ionospheric plasma structures from all-sky airglow images using deep learning framework. *Advances in Space Research*, *76*(4), 2302–2314. https://doi.org/10.1016/j.asr.2025.05.086

Charonko, J. J., & Vlachos, P. P. (2013). Estimation of uncertainty bounds for individual particle image velocimetry measurements from cross-correlation peak ratio. *Measurement Science and Technology*, *24*(6), 065301.

Dean, R. B., & Dixon, W. J. (1951). Simplified Statistics for Small Numbers of Observations. *Analytical Chemistry*, *23*(4), 636–638. https://doi.org/10.1021/ac60052a025

Dutta, A., & Zisserman, A. (2019). The VIA Annotation Software for Images, Audio and Video. In *Proceedings of the 27th ACM International Conference on Multimedia* (pp. 2276–2279). Nice France: ACM. https://doi.org/10.1145/3343031.3350535

Garcia, F. J., Kelley, M. C., Makela, J. J., Sultan, P. J., Pi, X., & Musman, S. (2000). Mesoscale structure of the midlatitude ionosphere during high geomagnetic activity: Airglow and GPS observations. Journal of Geophysical Research: Space Physics, 105(A8), 18417-18427.

Guerra, M., Cesaroni, C., Fiorentino, N., Tosone, F., Ventriglia, V., Pica, E., et al. (2025). Automatic detection of large-scale traveling ionospheric disturbances using GNSS data and image processing techniques. Space Weather, 23, e2025SW004423. https://doi.org/10.1029/2025SW004423

Hocke, K., & Schlegel, K. (1996). A review of atmospheric gravity waves and travelling ionospheric disturbances: 1982–1995. In Annales geophysicae (Vol. 14, No. 9, p. 917).

Hozumi, Y., Saito, A., Nishioka, M., Sakanoi, T., Yue, J., Chou, M.-Y., et al. (2025). Medium-Scale Traveling Ionospheric Disturbances Observed by Nadir-Viewing 630 nm Airglow Imaging From the International Space Station. *Journal of Geophysical Research: Space Physics*, *130*(10), e2025JA034097. https://doi.org/10.1029/2025JA034097

Huang, F., Lei, J., Dou, X., Luan, X., & Zhong, J. (2018). Nighttime Medium-Scale Traveling Ionospheric Disturbances From Airglow Imager and Global Navigation Satellite Systems Observations. *Geophysical Research Letters*, *45*(1), 31–38. https://doi.org/10.1002/2017GL076408

Huang, F., Lei, J., Otsuka, Y., Luan, X., Liu, Y., Zhong, J., & Dou, X. (2021). Characteristics of Medium-Scale Traveling Ionospheric Disturbances and Ionospheric Irregularities at Mid-Latitudes Revealed by the Total Electron Content Associated With the Beidou Geostationary Satellite. *IEEE Transactions on Geoscience and Remote Sensing*, *59*(8), 6424–6430. https://doi.org/10.1109/TGRS.2020.3032741

Jocher, G., Chaurasia, A., & Qiu, J. (2023). Ultralytics YOLOv8 (Version 8.0.0). Retrieved from https://github.com/ultralytics/ultralytics

Kale, K., Pawar, S., & Dhulekar, P. (2015). Moving object tracking using optical flow and motion vector estimation. In *2015 4th International Conference on Reliability, Infocom Technologies and Optimization (ICRITO) (Trends and Future Directions)* (pp. 1–6). https://doi.org/10.1109/ICRITO.2015.7359323

Kapil, C., & Seemala, G. K. (2024). Machine learning approach for detection of plasma depletions from TEC. Advances in Space Research, 73(7), 3833-3844. https://doi.org/10.1016/j.asr.2023.04.042

Lai, C., Xu, J., Lin, Z., Wu, K., Zhang, D., Li, Q., et al. (2023). Statistical Characteristics of Nighttime Medium-Scale Traveling Ionospheric Disturbances From 10-Years of Airglow Observation by the Machine Learning Method. *Space Weather*, *21*(5). https://doi.org/10.1029/2023SW003430

Le Moing, G., Ponce, J., & Schmid, C. (2024). Dense Optical Tracking: Connecting the Dots (pp. 19187–19197). Presented at the Proceedings of the IEEE/CVF Conference on Computer Vision and Pattern Recognition. Retrieved from https://openaccess.thecvf.com/content/CVPR2024/html/Le_Moing_Dense_Optical_Tracking_Connecting_the_Dots_CVPR_2024_paper.html

Le, X., Ren, X., Mei, D., Hu, F., Shinbori, A., Nishioka, M., et al. (2025). Intelligent Detection and Propagation Parameter Calculation of Medium-Scale Traveling Ionospheric Disturbances Based on YOLO and Feature Matching. *IEEE Transactions on Geoscience and Remote Sensing*, 1–1. https://doi.org/10.1109/TGRS.2025.3613933

Leonida, K. L., Sevilla, K. V., & Manlises, C. O. (2022). A Motion-Based Tracking System Using the Lucas-Kanade Optical Flow Method. In *2022 14th International Conference on Computer and Automation Engineering (ICCAE)* (pp. 86–90). https://doi.org/10.1109/ICCAE55086.2022.9762423

Liu, Y., Zhou, C., Xu, T., Tang, Q., Deng, Z. X., Chen, G. Y., and Wang, Z. K. (2021). Review of ionospheric irregularities and ionospheric electrodynamic coupling in the middle latitude region. Earth Planet. Phys., 5(5), 462–482. DOI: 10.26464/epp2021025

Lucas, B. D., & Kanade, T. (1981). An Iterative Image Registration Technique with an Application to Stereo Vision. In *IJCAI'81: 7th international joint conference on Artificial intelligence* (Vol. 2, pp. 674–679). Vancouver, Canada. Retrieved from https://hal.science/hal-03697340

Maeda, J., & Heki, K. (2015). Morphology and dynamics of daytime mid-latitude sporadic-E patches revealed by GPS total electron content observations in Japan. *Earth, Planets and Space*, *67*(1), 89. https://doi.org/10.1186/s40623-015-0257-4

Makela, J. J. (2006). A review of imaging low-latitude ionospheric irregularity processes. *Journal of Atmospheric and Solar-Terrestrial Physics*, *68*(13), 1441–1458. https://doi.org/10.1016/j.jastp.2005.04.014

Mondal, S., Srivastava, A., Yadav, V., Sarkhel, S., Sunil Krishna, M. V., Rao, Y. K., & Singh, V. (2019). Allsky airglow imaging observations from Hanle, Leh Ladakh, India: Image analyses and first results. *Advances in Space Research*, *64*(10), 1926–1939. https://doi.org/10.1016/j.asr.2019.05.047

Mutasov, G., Supnithi, P., Budtho, J., Tongkasem, N., Nishioka, M., Perwitasari, S., & Myint, L. M. M. (2025). Classification of Equatorial Ionospheric Irregularities Using Unsupervised Machine Learning Based on Spatiotemporal ROTI Keograms. IEEE Journal of Selected Topics in Applied Earth Observations and Remote Sensing.

Naito, H., Shiokawa, K., Otsuka, Y., Fujinami, H., Tsuboi, T., Sakanoi, T., et al. (2022). Three-Dimensional Fourier Analysis of Atmospheric Gravity Waves and Medium-Scale Traveling Ionospheric Disturbances Observed in Airglow Images in Hawaii Over Three Years. *Journal of Geophysical Research: Space Physics*, *127*(10), e2022JA030346. https://doi.org/10.1029/2022JA030346

Okoh, D., Cesaroni, C., Rabiu, B., Shiokawa, K., Otsuka, Y., Ogunjo, S., et al. (2025). A Bootstrapping Convolutional Neural Network Technique for Optimizing Automated Detection of Equatorial Plasma Bubbles by Optical All-Sky Imagers. *Earth and Space Science*, *12*(6), e2024EA004117. https://doi.org/10.1029/2024EA004117

Otsuka, Y., K. Shiokawa, and T. Ogawa (2012), Disappearance of equatorial plasma bubble after interaction with mid-latitude medium-scale traveling ionospheric disturbance, Geophys. Res. Lett., 39, L14105, doi:10.1029/2012GL052286.

Otsuka, Y., Shiokawa, K., Ogawa, T., & Wilkinson, P. (2004). Geomagnetic conjugate observations of medium-scale traveling ionospheric disturbances at midlatitude using all-sky airglow imagers. *Geophysical Research Letters*, *31*(15). https://doi.org/10.1029/2004GL020262

Otsuka, Yuichi. (2021). Medium-Scale Traveling Ionospheric Disturbances. In *Ionosphere Dynamics and Applications* (pp. 421–437). wiley. https://doi.org/10.1002/9781119815617.ch18

Otsuka, Yuichi, Shinbori, A., Tsugawa, T., & Nishioka, M. (2021). Solar activity dependence of medium-scale traveling ionospheric disturbances using GPS receivers in Japan. *Earth, Planets and Space*, *73*(1). https://doi.org/10.1186/s40623-020-01353-5

Padfield, D. (2012). Masked Object Registration in the Fourier Domain. *IEEE Transactions on Image Processing*, *21*(5), 2706–2718. https://doi.org/10.1109/TIP.2011.2181402

Patgiri, D., Rathi, R., Yadav, V., Chakrabarty, D., Sunil Krishna, M. V., Kannaujiya, S., et al. (2024a). A Rare Simultaneous Detection of a Mid-Latitude Plasma Depleted Structure in O($^1$D) 630.0 and O($^1$S) 557.7 nm All-Sky Airglow Images on a Geomagnetically Quiet Night. *Geophysical Research Letters*, *51*(14). https://doi.org/10.1029/2023GL106900

Patgiri, D., Nirwal, S., Rathi, R., Yadav, V., Sunil Krishna, M. V., Kannaujiya, S., et al. (2024b). Role of Coupled Electrodynamics of E- and F-Regions on the Rapid Poleward Extension of EMSTID Fronts. *Journal of Geophysical Research: Space Physics*, *129*(12), e2024JA032991. https://doi.org/10.1029/2024JA032991

Patgiri, Dipjyoti, Rathi, R., Yadav, V., Sarkhel, S., Chakrabarty, D., Mondal, S., et al. (2024c). A case study on multiple self-interactions of MSTID bands: New insights. *Advances in Space Research*, *73*(7), 3595–3612. https://doi.org/10.1016/j.asr.2023.05.047

Patgiri, Dipjyoti, Otsuka, Y., Shreedevi, P. R., Kakoti, G., Ranjan, A. K., Rathi, R., et al. (2025). Unusual Phase Alignment of Nighttime Electrified Medium-Scale Traveling Ionospheric Disturbance After Moderate Geomagnetic Storms and Response to Substorms. *Geophysical Research Letters*, *52*(20), e2025GL116392. https://doi.org/10.1029/2025GL116392

Rathi, R., Yadav, V., Mondal, S., Sarkhel, S., Krishna, M. V. S., & Upadhayaya, A. K. (2021). Evidence for simultaneous occurrence of periodic and single dark band MSTIDs over geomagnetic low-mid latitude transition region. *Journal of Atmospheric and Solar-Terrestrial Physics*, *215*. https://doi.org/10.1016/j.jastp.2021.105588

Rathi, R., Yadav, V., Mondal, S., Sarkhel, S., Sunil Krishna, M. V., Upadhayaya, A. K., et al. (2022). A Case Study on the Interaction Between MSTIDs' Fronts, Their Dissipation, and a Curious Case of MSTID's Rotation Over Geomagnetic Low-Mid Latitude Transition Region. *Journal of Geophysical Research: Space Physics*, *127*(4). https://doi.org/10.1029/2021JA029872

Rathi, R., Sivakandan, M., Chakrabarty, D., Sunil Krishna, M. V., Upadhayaya, A. K., & Sarkhel, S. (2025). Evidence for the evolution and decay of an electrified medium scale traveling ionospheric disturbances during two consecutive substorms: First results. *Advances in Space Research*, *75*(6), 4896–4909. https://doi.org/10.1016/j.asr.2025.01.007

Rathi, Rahul, Gurram, P., Mondal, S., Yadav, V., Sarkhel, S., Sunil Krishna, M. V., & Upadhayaya, A. K. (2023). Unusual simultaneous manifestation of three non-interacting mid-latitude ionospheric plasma structures. *Advances in Space Research*. https://doi.org/10.1016/j.asr.2023.04.038

Redmon, J., Divvala, S., Girshick, R., & Farhadi, A. (2016). You Only Look Once: Unified, Real-Time Object Detection. In *2016 IEEE Conference on Computer Vision and Pattern Recognition (CVPR)* (pp. 779–788). Las Vegas, NV, USA: IEEE. https://doi.org/10.1109/CVPR.2016.91

Rorabacher, D. B. (1991). Statistical treatment for rejection of deviant values: critical values of Dixon's "Q" parameter and related subrange ratios at the 95% confidence level. *Analytical Chemistry*, *63*(2), 139–146. https://doi.org/10.1021/ac00002a010

Sedlak, R., Welscher, A., Hannawald, P., Wüst, S., Lienhart, R., & Bittner, M. (2023). Analysis of 2D airglow imager data with respect to dynamics using machine learning. *Atmospheric Measurement Techniques*, *16*(12), 3141–3153. https://doi.org/10.5194/amt-16-3141-2023

Shiokawa, K., Otsuka, Y., Ihara, C., Ogawa, T., & Rich, F. J. (2003). Ground and satellite observations of nighttime medium-scale traveling ionospheric disturbance at midlatitude. *Journal of Geophysical Research: Space Physics*, *108*(A4). https://doi.org/10.1029/2002JA009639

Shiokawa, K., Ihara, C., Otsuka, Y., & Ogawa, T. (2003). Statistical study of nighttime medium-scale traveling ionospheric disturbances using midlatitude airglow images. *Journal of Geophysical Research: Space Physics*, *108*(A1). https://doi.org/10.1029/2002JA009491

Shorten, C., & Khoshgoftaar, T. M. (2019). A survey on Image Data Augmentation for Deep Learning. *Journal of Big Data*, *6*(1), 60. https://doi.org/10.1186/s40537-019-0197-0

Sivakandan, M., Chakrabarty, D., Ramkumar, T. K., Guharay, A., Taori, A., & Parihar, N. (2019). Evidence for Deep Ingression of the Midlatitude MSTID Into As Low as 3.5° Magnetic Latitude. *Journal of Geophysical Research: Space Physics*, *124*(1), 749–764. https://doi.org/10.1029/2018JA026103

Sivakandan, M., Mondal, S., Sarkhel, S., Chakrabarty, D., Sunil Krishna, M. V., Chaitanya, P. P., et al. (2020). Mid-Latitude Spread-F Structures Over the Geomagnetic Low-Mid Latitude Transition Region: An Observational Evidence. *Journal of Geophysical Research: Space Physics*, *125*(5). https://doi.org/10.1029/2019JA027531

Sivakandan, M., Mondal, S., Sarkhel, S., Chakrabarty, D., Sunil Krishna, M. V., Upadhayaya, A. K., et al. (2021). Evidence for the In-Situ Generation of Plasma Depletion Structures Over the Transition Region of Geomagnetic Low-Mid Latitude. *Journal of Geophysical Research: Space Physics*, *126*(9). https://doi.org/10.1029/2020JA028837

Sobral, J. H. A., Takahashi, H., Abdu, M. A., Muralikrishna, P., Sahai, Y., & Zamlutti, C. J. (1992). O(1S) and O(1D) quantum yields from rocket measurements of electron densities and 557.7 and 630.0 nm emissions in the nocturnal F-region. *Planetary and Space Science*, *40*(5), 607–619. https://doi.org/10.1016/0032-0633(92)90002-6

Sun, D., Roth, S., & Black, M. J. (2014). A Quantitative Analysis of Current Practices in Optical Flow Estimation and the Principles Behind Them. *International Journal of Computer Vision*, *106*(2), 115–137. https://doi.org/10.1007/s11263-013-0644-x

Sun, L., J. Xu, W. Wang, X. Yue, W. Yuan, B. Ning, D. Zhang, and F. C. de Meneses (2015), Mesoscale field-aligned irregularity structures (FAIs) of airglow associated with medium-scale traveling ionospheric disturbances (MSTIDs), J. Geophys. Res. Space Physics, 120, 9839–9858, doi:10.1002/2014JA020944.

Takeo, D., Shiokawa, K., Fujinami, H., Otsuka, Y., Matsuda, T. S., Ejiri, M. K., et al. (2017). Sixteen year variation of horizontal phase velocity and propagation direction of mesospheric and thermospheric waves in airglow images at Shigaraki, Japan. *Journal of Geophysical Research: Space Physics*, *122*(8), 8770–8780. https://doi.org/10.1002/2017JA023919

Tsuboi, T., Shiokawa, K., Otsuka, Y., Fujinami, H., & Nakamura, T. (2023). Statistical Analysis of the Horizontal Phase Velocity Distribution of Atmospheric Gravity Waves and Medium-Scale Traveling Ionospheric Disturbances in Airglow Images Over Sata (31.0°N, 130.7°E), Japan. *Journal of Geophysical Research: Space Physics*, *128*(12). https://doi.org/10.1029/2023JA031600

Tsuchiya, S., Shiokawa, K., Otsuka, Y., Nakamura, T., Yamamoto, M., Connors, M., et al. (2020). Wavenumber Spectra of Atmospheric Gravity Waves and Medium-Scale Traveling Ionospheric Disturbances Based on More Than 10-Year Airglow Images in Japan, Russia, and Canada. *Journal of Geophysical Research: Space Physics*, *125*(3), e2019JA026807. https://doi.org/10.1029/2019JA026807

Westerweel, J. (1997). Fundamentals of digital particle image velocimetry. *Measurement Science and Technology*, *8*(12), 1379. https://doi.org/10.1088/0957-0233/8/12/002

Wu, K., Xu, J., Wang, W., Sun, L., & Yuan, W. (2021). Interaction of oppositely traveling medium-scale traveling ionospheric disturbances observed in low latitudes during geomagnetically quiet nighttime. Journal of Geophysical Research: Space Physics, 126, e2020JA028723. https://doi.org/10.1029/2020JA028723

Yadav, V., Rathi, R., Sarkhel, S., Chakrabarty, D., Sunil Krishna, M. V., & Upadhayaya, A. K. (2021a). A Unique Case of Complex Interaction Between MSTIDs and Mid-Latitude Field-Aligned Plasma Depletions Over Geomagnetic Low-Mid Latitude Transition Region. *Journal of Geophysical Research: Space Physics*, *126*(1). https://doi.org/10.1029/2020JA028620

Yadav, V., Rathi, R., Gaur, G., Sarkhel, S., Chakrabarty, D., Sunil Krishna, M. V., et al. (2021b). Interaction between nighttime MSTID and mid-latitude field-aligned plasma depletion structure over the transition region of geomagnetic low-mid latitude: First results from Hanle, India. *Journal of Atmospheric and Solar-Terrestrial Physics*, *217*, 105589. https://doi.org/10.1016/j.jastp.2021.105589

Zhang, H., Grießbach, D., Wohlfeil, J., & Börner, A. (2018). Uncertainty Model for Template Feature Matching. In M. Paul, C. Hitoshi, & Q. Huang (Eds.), *Image and Video Technology* (Vol. 10749, pp. 406–420). Cham: Springer International Publishing. https://doi.org/10.1007/978-3-319-75786-5_33

Zhang, T., Luo, Y., Zhang, B., Jiang, F., Wang, T., Xiao, S., et al. (2026). Automatic detection of equatorial plasma bubbles using deep neural networks. *Journal of Atmospheric and Solar-Terrestrial Physics*, *278*, 106687. https://doi.org/10.1016/j.jastp.2025.106687

Zhong, J., Zou, Z., Wu, K., Xu, J., Lu, Y., Sun, L., et al. (2025). Automatic Detection and Feature Extraction of Equatorial Plasma Bubbles From All-Sky Airglow Image Based on Machine Learning. *Space Weather*, *23*(5), e2025SW004336. https://doi.org/10.1029/2025SW004336